\documentclass[useAMS]{mn2e}
\usepackage{graphicx}
\usepackage[authoryear]{natbib}

\title[GRB afterglows as they come into view of Swift UVOT]{A study of gamma-ray burst afterglows as they first come into view of the Swift Ultraviolet and Optical Telescope}

\author[M.J. Page et al.]
{M.J. Page$^{1}$, 
S.R. Oates$^{2,1}$,
M. De Pasquale$^{3,1}$,
A.A. Breeveld$^{1}$, 
S.W.K. Emery$^{1}$,
\and 
N.P.M. Kuin$^{1}$,
F.E. Marshall$^{4}$,
M.H. Siegel$^{5}$,
P.W.A. Roming$^{6}$\\ 
\\
$^{1}$Mullard Space Science Laboratory, University College London,
Holmbury St Mary, Dorking, Surrey, RH5 6NT, UK\\
$^{2}$Department of Physics, University of Warwick, Coventry CV4 7AL, UK\\
$^{3}$Department of Astronomy and Space Sciences, Istanbul University, Turkey\\
$^{4}$Astrophysics Science Division, NASA Goddard Space Flight Centre, Greenbelt MD, 20771 USA\\
$^{5}$Department of Astronomy and Astrophysics, The Pennsylvania State University, University Park, PA 16802, USA\\
$^{6}$Southwest Research Institute Space Science and Engineering Division, 6220 Culebra Road San Antonio, TX 78238-5166, USA
}

\begin{document}

\date{Accepted ----. Received ----; in original form ----}

\pagerange{\pageref{firstpage}--\pageref{lastpage}} 
\pubyear{2017}
\maketitle

\label{firstpage}

\begin{abstract}

  We examine the the emission from optically bright gamma-ray burst
  (GRB) afterglows as the Ultraviolet and Optical Telescope (UVOT) on
  the {\em Neil Gehrels Swift Observatory} first begins observing,
  following the slew to target the GRB, while the pointing of the {\em
    Swift} satellite is still settling.  We verify the photometric
  quality of the UVOT settling data using bright stars in the field of
  view.  
  In the majority of cases we find no problems with the
  settling exposure photometry, but in one case we excise the first
  second of the exposure to mitigate a spacecraft attitude
  reconstruction issue, and in a second case we exclude the first
  second of the exposure in which the UVOT photocathode voltage
  appears to be ramping up. 
  Of a sample of 23 afterglows which
  have peak V magnitudes $<16$, we find that all are detected in the
  settling exposures, when {\em Swift} arrives on target. For 9 of the
  GRBs the UVOT settling exposure took place before the conclusion of
  the prompt gamma-ray emission. Five of these GRBs have well defined
  optical peaks after the settling exposures, with rises of
  $>0.5$~mag in their optical lightcurves, and there is a marginal
  trend for these GRBs to have long $T_{90}$. Such a trend is expected
  for thick-shell afterglows, but the temporal indices of the optical
  rises and the timing of the optical peaks appear to rule out thick
  shells.
\end{abstract}


\section{Introduction}
\label{sec:introduction}

Gamma-ray bursts are the most powerful cosmic explosions. They are
observed as transient sources of gamma-rays, which typically last
between $10^{-2}$ and $10^{3}$~s \citep{kouveliotou94}. This gamma-ray
emission, which is usually referred to as the {\em prompt} gamma-ray
emission, is followed by a longer lasting afterglow from X-ray to
radio wavelenghts \citep[e.g.][]{depasquale06}.  The standard paradigm
for the prompt emission is that it arises from shocks associated with
the collisions of shells ejected with different Lorentz factors in a
highly relativistic explosion. The radiation generation mechanism(s)
for the prompt emission remain the subject of debate, with
possibilities including synchrotron emission from the shocks
\citep[e.g. ][]{rees94}, Compton upscattering of photospheric emission
\citep[e.g. ][]{rees05} or magnetic reconnection
\citep[e.g. ][]{zhang11}.  The emission in these models is often
described as internal emission because it takes place within the
relativistic ejecta, before the ejecta are slowed significantly
through interaction with the surrounding medium.  The afterglow is
thought to originate in the shocks arising from the collision of the
ejecta with the surrounding interstellar medium; these shocks are therefore
described as the external shocks \citep{rees92}.

Observational data are most constraining for models of the prompt emission and afterglow when the afterglow observations are obtained very quickly after, if not simultaneously with, the prompt emission. 
The {\em Neil Gehrels Swift Observatory} (hereafter {\em Swift}) carries a suite of instruments on a platform with rapid repointing capabilities and a significant degree of autonomy to permit the study of GRBs and their early afterglow emission \citep{gehrels04}.
{\em Swift} begins its afterglow observations most rapidly when it 
detects and localises GRBs
on-board with the Burst Alert Telescope \citep[BAT;
][]{barthelmy05}, prompting {\em Swift} to execute an automatic slew
to point the Ultraviolet and Optical Telescope \citep[UVOT;
][]{roming05} and X-ray Telescope \citep[XRT; ][]{burrows05} at the
target. Response times of 1-2 minutes are common.

During the slew, the UVOT is protected from damage by bright stars
passing through the field of view by maintaining the photocathode in a
low-voltage state. At the conclusion of the slew, {\em Swift} signals
to the UVOT that it has arrived at the target, and the UVOT begins its
first exposure. At this point, {\em Swift}'s reaction wheels are stabilising
the satellite, such that the pointing direction is in transition from
a motion of many arcseconds per second to a stable pointing. The
exposure taken by UVOT during this transition is known as the
settling exposure; 
its nominal duration is 10~s.
The settling exposure is almost always taken
through the $V$ filter; exceptions are found in a small period at the
beginning of the mission when the default settling filter was UVM2,
and occasions when the field of view of UVOT contains one or more
stars which are recorded in the on-board catalogue as being too bright
for safe observations through the $V$ filter.

Despite being the earliest UVOT exposure of {\em Swift}-discovered
GRBs, the settling exposure is frequently ignored or discarded
\citep[e.g. ][]{oates12}. The main reason is that photometry derived
from the settling exposure is regarded as uncertain because of the
changing UVOT photocathode voltage at the beginning of the exposure;
the rapidly changing spacecraft attitude can also be a cause for
concern \citep{roming17}. The settling exposure usually begins 10-15
seconds before the first settled exposure, and so represents a
significant shifting forward of the afterglow observations, if the
photometry can be trusted.

The detector of the UVOT is a micro-channel plate (MCP) intensified
charge coupled device (CCD) \citep[MIC;][]{fordham89}. It detects the
photons individually with a time resolution limited by the frame time
of the CCD, which is usually 11~ms. Data can be recorded and relayed
to the ground as event lists, in which the arrival times and positions
of the individual photons are retained, or as an image accumulated
over a time interval. To meet data and telemetry limits, most UVOT
data are recorded as images. Settling exposures are recorded as event
lists. Although the satellite is moving during the settling exposure,
the changing attitude of the satellite is taken into account when 
the photons are
assigned sky coordinates. Thus an image of the sky can be constructed
from the settling-exposure event list in which stars have shapes
consistent with the normal UVOT point spread function, and there is no
evidence of trailing.

In this paper, we verify the photometric quality of the settling
exposures of a group of relatively bright GRB afterglows observed with
{\em Swift} UVOT. Photometry of the GRB afterglows is then derived
from the settling exposures, and the implications of these very early
data are examined. In Section~\ref{sec:method} we describe the sample
selection, observations and data reduction. Our results are presented
in Section~\ref{sec:results} and discussed in
Section~\ref{sec:discussion}. Our conclusions are given in
Section~\ref{sec:conclusions}. An appendix provides the measurements
which were used to verify the photometric quality of the settling
exposures.  All magnitudes are given in the UVOT Vega system
\citep{poole08}. All GRB times are referenced to the beginning of $T_{90}$, 
which is the time period containing 90 per cent of the gamma-ray emission, 
as recorded in the {\it Swift} BAT on-line 
processing\footnote{http://gcn.gsfc.nasa.gov/swift\_gnd\_ana.html}. 

\begin{figure}
\begin{center}
\includegraphics[width=75mm, angle=0]{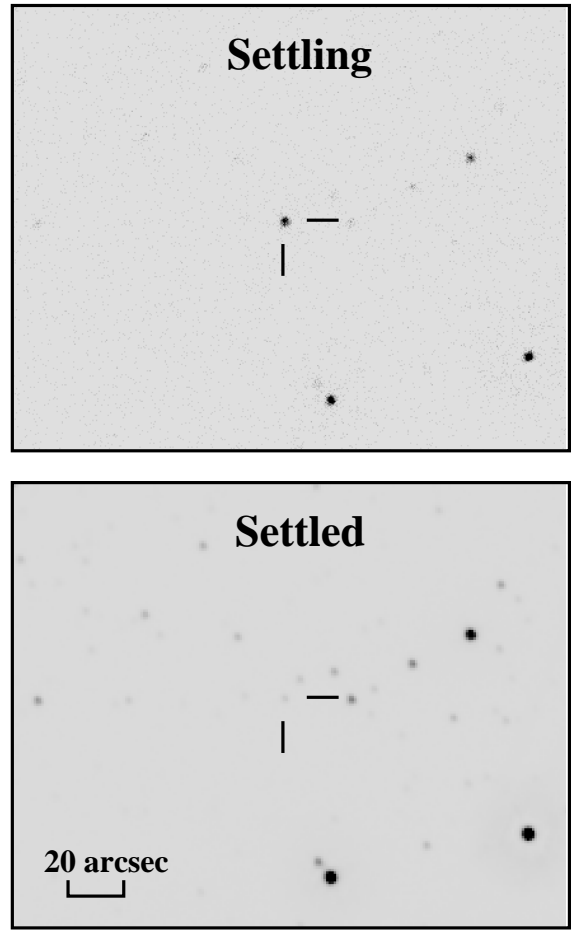}
\caption{Examples of settling (top) and settled (bottom) images 
for GRB\,050525A, constructed via the procedure described 
in Section~\ref{sec:observations}. Note that only a fraction of 
the UVOT field of view is shown. The GRB afterglow is marked at the centre.
The pixel size for the settling image is 0.5~arcsec $\times$ 0.5~arcsec, because the UVOT event list retains the full spatial sampling of the UVOT detector array, while the settled image has 1~arcsec $\times$ 1~arcsec pixels, reflecting the on-board $2\times 2$ binning usually employed for image-mode data.}
\label{fig:example}
\end{center}
\end{figure}

\section{Sample selection, observations and data reduction}
\label{sec:method}

\subsection{Sample Selection}

Our objective is to examine the early periods of gamma-ray burst
optical afterglow light curves to determine what fraction of the
sample shows significant afterglow emission when {\em Swift} UVOT
first arrives on target. Therefore we require a sample of gamma-ray
burst afterglows that (a) are optically bright enough that we can
expect to detect the afterglow in settling observations at a
significant fraction of the peak flux, (b) were observed soon after
the trigger and (c) have redshifts so that their observed properties
can be transformed to the rest-frame. 

With these criteria in mind, we have started with the sample of GRBs
studied by \citet{oates12}. The sample of \citet{oates12} was drawn
from the second UVOT GRB catalogue \citep{roming17}, which contains
all of the GRBs observed with {\em Swift} UVOT from the launch of {\em
  Swift} until 25 December 2010. For inclusion in their sample,
\citet{oates12} required that a GRB must have been observed by UVOT
within the first 400s after the BAT trigger and have a measurement of
redshift, reflecting our criteria (b) and (c). For a typical settling
exposure duration of 10~s, the limiting magnitude for a measurement
with a signal to noise of 3 is $V=17.5$~mag. Our criterion (a) must
therefore be more stringent than the peak brightness limit ($V \leq
17.89$~mag) adopted by \citet{oates12}. For our study we have included
only objects which have a peak $V$ magnitude of $\leq 16.0$, which
corresponds to a sample of 23 GRBs that are listed in Table~\ref{tab:grbs}.

\begin{table*}
\caption{The GRB sample. $T_{\rm settle}$ is mid-time of the settling exposure referenced to the beginning of $T_{90}$. Redshifts and brightest $V$ magnitudes are those used by \citet{oates12}, except for GRB\,080319B, for which the UVOT data were saturated at early times. The brightest $V$ magnitude listed for GRB\,080319B is the brightest $V$ magnitude from subsequent UVOT observations, obtained by read-out-streak photometry \citep{page13}. The settling $V$ magnitude for GRB\,080319B listed here is the magnitude recorded by TORTORA at the time of the UVOT settling exposure \citep{racusin08}.
}
\label{tab:grbs}
\begin{tabular}{lcccrr}
GRB         &Redshift&Brightest $V$ mag&Settling $V$ mag&$T_{\rm settle}$&$T_{90}$\\
            &        &post-settling    &(mag)           &(s)        &(s)     \\
            &        &(mag)            &                &           &        \\
\hline
&&&&&\\
GRB\,050525A&0.606&$13.47\pm0.06$&$13.42\pm0.06$& 70.2&  8.8\\
GRB\,050801 &1.38 &$15.18\pm0.12$&$15.49\pm0.14$& 56.3& 19.4\\
GRB\,050922C&2.198&$14.46\pm0.06$&$13.92\pm0.06$&104.3&  4.5\\
GRB\,060418 &1.489&$14.58\pm0.04$&$15.96\pm0.17$&136.7&144.0\\
GRB\,060607 &3.43 &$14.49\pm0.04$&$16.62\pm0.26$&103.5& 60.5\\
GRB\,060908 &2.43 &$14.97\pm0.05$&$14.70\pm0.09$& 75.0& 19.3\\
GRB\,061007 &1.261&$11.93\pm0.03$&$<11.67$      & 72.0& 75.3\\
GRB\,061021 &0.77 &$15.46\pm0.05$&$15.00\pm0.11$& 63.9& 46.2\\
GRB\,061121 &1.314&$15.41\pm0.05$&$17.52\pm0.43$& 39.5& 81.2\\
GRB\,070318 &0.836&$15.34\pm0.05$&$16.18\pm0.20$& 58.8&108.0\\
GRB\,080319B&0.937&$10.07\pm0.24$&$ 6.43\pm0.06$& 51.0& 45.1\\
GRB\,080721 &2.602&$13.41\pm0.04$&$12.99\pm0.06$&118.5& 64.0\\
GRB\,080810 &3.35 &$13.58\pm0.04$&$13.49\pm0.06$& 77.1&108.8\\
GRB\,081008 &1.967&$14.29\pm0.03$&$14.37\pm0.08$&125.6&185.5\\
GRB\,081203A&2.05 &$13.04\pm0.02$&$15.87\pm0.17$& 93.4&221.0\\
GRB\,081222 &2.77 &$14.93\pm0.05$&$12.80\pm0.06$& 46.5& 33.0\\
GRB\,090401B&3.1  &$15.22\pm0.06$&$14.90\pm0.11$& 67.8&186.5\\
GRB\,090424 &0.544&$14.49\pm0.04$&$13.74\pm0.07$& 79.7& 49.5\\
GRB\,090618 &0.54 &$14.30\pm0.04$&$13.97\pm0.07$&105.9&113.2\\
GRB\,090812 &2.452&$15.66\pm0.06$&$15.70\pm0.15$& 73.7& 66.7\\
GRB\,091018 &0.971&$14.38\pm0.06$&$13.77\pm0.07$& 56.0&  4.4\\
GRB\,091020 &1.71 &$15.08\pm0.04$&$14.86\pm0.11$& 80.2& 39.0\\
GRB\,100906A&1.727&$14.50\pm0.04$&$14.11\pm0.08$& 73.7& 14.3\\
&&&&&\\
\hline
\end{tabular}
\end{table*}

\subsection{Observations, data reduction and analysis}
\label{sec:observations}

All of the GRBs in the sample were observed with {\em Swift} UVOT in
Automated Target (AT) mode, and the analysis is restricted to those
observations obtained in the initial observing segment, roughly the
first 24 hours from the detection of the burst.

For each GRB the UVOT $V$-band sky images and event data were
retrieved from the UK {\em Swift} Science Data
Centre\footnote{http://www.swift.ac.uk}. Images were screened visually
for problems such as source trailing and any affected images were
discarded. Then, for each GRB all remaining images except for that of
the settling exposure were summed using the standard {\em Swift} {\sc
  ftool} task {\sc uvotimsum} to produce a single settled image.  For
the settling exposures, sky images were constructed from the event
lists using {\sc evselect}. The resulting sky images were inspected
for attitude problems indicated by trailing of stars. Only in the case
of GRB\,061007 was a problem identified; in this case the problem was
solved by discarding the first second of the settling exposure. Then,
for each GRB the settling image and settled image were registered and
aspect corrected to the same astrometric reference frame using the
USNO-A2.0 source catalogue \citep{monet98}. 
Examples of the settling and settled images constructed 
in this way are shown in Fig.~\ref{fig:example}.

Photometric measurements of between one and 3 reference stars in the
settling and settled images of each GRB field were then obtained to
assess the photometric properties of the settling images. The
brightest non-saturated stars in each field were chosen as the
reference stars in order to maximise the accuracy of the
photometry. Circular apertures of 5~arcsec radius were employed for
the source measurements, as is standard practice for measurement of
bright sources in UVOT \citep{poole08}, and background was measured
from a source-free circular aperture of 25~arcsec radius. The
measurements were carried out with the standard {\em Swift } {\sc
  ftool} task {\sc uvotmaghist}, which corrects for coincidence-loss,
the evolution of the UVOT sensitivity with time, and the large scale
sensitivity variations over the detector \citep[see ][]{breeveld10}.

Photometric measurements of the GRB afterglows in the settling
exposures were carried out by following the same procedure as for the
reference stars.

\section{Results}
\label{sec:results}

\begin{figure}
\begin{center}
\includegraphics[width=60mm, angle=270]{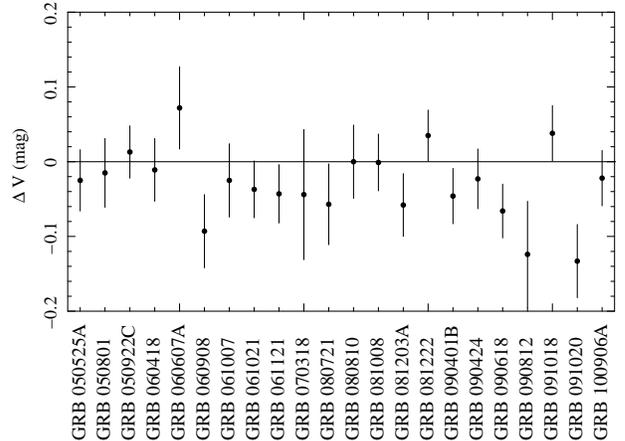}
\caption{Weighted mean difference $\Delta V$ between the magnitudes of the reference stars in the settling and settled exposures for each field. $\Delta V$ is computed as $V_{\rm settled}-V_{\rm settling}$ where $V_{\rm settling}$ is the $V$ magnitude in the settling exposure and $V_{\rm settled}$ is the $V$ magnitude in the subsequent settled exposures, so that negative values correspond to the reference stars appearing fainter in the settling image.}
\label{fig:fidelity}
\end{center}
\end{figure}

Fig.~\ref{fig:fidelity} shows the photometric offsets of the settling
exposures with respect to the later, settled exposures as measured
with the reference stars in each field. Further details of the reference-star 
measurements are given in Appendix A. The majority of the offsets
are negative, indicating that the reference stars appear to be fainter
in the settling exposures, but most offsets are small (all bar two are
below 0.1 magnitude) and only one, corresponding to GRB~091020, is
individually significant at $>2 \sigma$. 
Where a bright enough comparison star is available, we have further investigated the photometric stability of the settling exposures by splitting them into 1~s chunks and generating lightcurves for the comparison stars; the lightcurves are given in Appendix A. In only one case do we see evidence that the ramp-up of the photocathode voltage impinges on the settling exposure: the comparison star in the field of GRB~091020 appears a magnitude fainter in the first second than in the remainder of the settling exposure. This problem can be remedied by excluding the first second of the settling exposure from our analysis. So doing, the photometric offset of the GRB~091020 field changes from $-0.13\pm0.05$ to $-0.05\pm0.05$, consistent with zero at 1~$\sigma$.
After this correction to the GRB~091020 field, the weighted mean offset for
all GRB fields is $-0.021 \pm 0.009$ mag. This mean offset is sufficiently marginal (2~$\sigma$) that we cannot distinguish whether it is a genuine instrumental effect or a statistical fluctuation. The size of the mean offset is sufficiently small (less than half of the 1~$\sigma$ uncertainty on the best settling-exposure photometry in our GRB sample) that we do not consider any correction to the settling-exposure photometry to be necessary.

Photometry for the GRBs from the settling exposures are given in
Table~\ref{tab:grbs}, together with their redshifts, some optical
measurements at later times from the lightcurves constructed by
\cite{oates12}, and the $T_{90}$ durations of their gamma-ray emission from the
on-line BAT
processing$^{1}$.
All of the GRBs were detected in the settling
exposure with a significance of $>3\sigma$ with respect to the Poission fluctuations in the background. 
Note that in the case of GRB\,061007, the count rate of the afterglow exceeds the calibrated limit for coincidence-loss correction in UVOT \citep[see ][]{poole08,page13} during the settling exposure. Consequently, we report an upper limit for the magnitude, corresponding to the maximum calibrated count rate, in Table~\ref{tab:grbs}. In statistical terms, GRB\,061007 is securely (3$\sigma$) brighter than this limit, using the statistical uncertainties derived by \citet{kuin08}. 

\begin{figure}
\begin{center}
\includegraphics[width=60mm, angle=270]{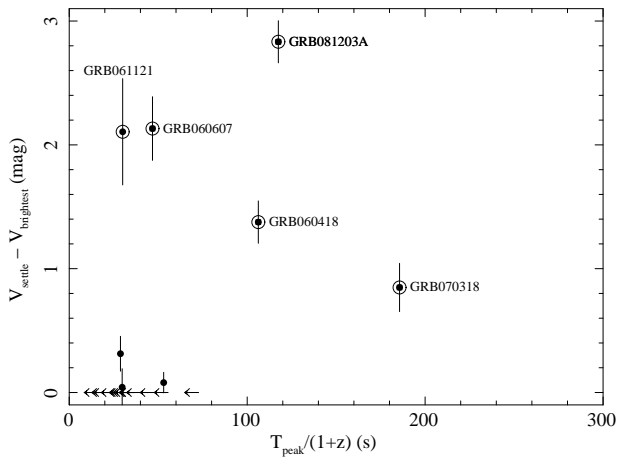}
\caption{Difference between the brightest magnitude and the settling
  magnitude for each GRB against the time at which the optical
  emission peaks, in the rest frame of the GRB. In the majority of
  cases, the settling exposure records the brightest magnitude 
  (i.e. $V_{\rm settle}=V_{\rm brightest}$), hence
  we have only an upper limit on the time at which the optical
  emission peaks. GRBs which brighten by more than 0.5 magnitudes
  after the settling exposure are indicated with large circles and
  are labelled individually.}
\label{fig:deltam_vs_tpeak}
\end{center}
\end{figure}

Fig.~\ref{fig:deltam_vs_tpeak} shows the difference between the
magnitude recorded in the settling exposure ($V_{\rm settle}$) and the
brightest magnitude recorded either in the settling exposure or the
subsequent UVOT lightcurve ($V_{\rm brightest}$) against the time of the
peak optical emission ($T_{\rm peak}$) in the rest frame of the GRB. 
When the brightest magnitude for a
GRB is recorded in the settling exposure, $T_{\rm peak}$ is shown as an
upper limit. The majority of the data points are clustered at the
bottom left of the figure. For these sources the settling exposure is
either the brightest measurement or differs by less than 0.4 mag from 
the brightest measurement. Away from the
bottom left corner are five GRBs which brighten significantly (by
0.5 -- 3 mag) after the settling exposure, and therefore have optical
lightcurves which show a well defined peak.

\begin{figure}
\begin{center}
\includegraphics[width=85mm]{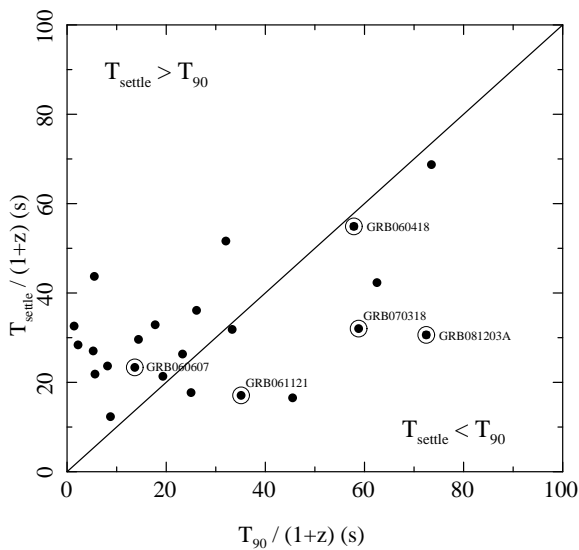}
\caption{Time $T_{\rm settle}$ at which the settling exposure was taken against $T_{90}$ of the gamma-ray emission. Both times have been converted to the rest frame of the GRB. The solid line indicates equality between $T_{\rm settle}$ and $T_{90}$. GRBs below the solid line were first observed with UVOT during the $T_{90}$ gamma-ray emitting period, while those above the line were only observed with UVOT after the end of $T_{90}$. GRBs for which a significant rise in the optical emission is observed are indicated with large circles and labelled individually as in Fig.~\ref{fig:deltam_vs_tpeak}. 
}
\label{fig:tsettle_vs_t90}
\end{center}
\end{figure}

Fig.~\ref{fig:tsettle_vs_t90} shows the mid-time of the settling
exposure $T_{\rm settle}$ against the gamma-ray duration $T_{90}$, with
both times transformed to the rest-frame of the GRB. The solid line
indicates $T_{\rm settle}=T_{90}$. Nine out of 23 points lie below the
line, implying that UVOT took its first exposures while the
prompt gamma-ray emission was still underway.

\section{Discussion}
\label{sec:discussion}

For our sample of 23 GRBs with bright optical afterglows, the optical
emission had already begun by the time of the {\em Swift} UVOT
settling exposure in every single case. In 9 cases, the settling
exposure took place before the end of the $T_{90}$ period in which
most of the prompt gamma rays are emitted. We do not observe a delay
between the end of the gamma-ray emission and the beginning of that in
the optical in any GRB.  Such a statement could be made on the basis
of some previous studies of GRB samples with rapid optical follow up
\citep[e.g.][]{rykoff09, oates09}, but the homogeneity and size of our
sample place this conclusion on a firmer statistical footing. 

Our study also provides a useful statistical context for studies of
gamma-ray bursts which show peaks in their optical lightcurves
\citep[e.g.][]{molinari07, panaitescu11, panaitescu13}. At the
peak-magnitude limit ($V=16.0$) of our study, only 5 out of 23 GRBs
show well-defined optical peaks, or $20\pm10$ per cent, 
within the time-frame of the observations. The need for even faster
optical response times for GRBs is brought into sharp focus: while a
significant anti-correlation has been observed between the timing and
brightness of the optical peak by \citet{panaitescu11} when such a
peak is observed, for the majority of GRBs we have only upper and
lower limits for these two quantities respectively, because our
observations begin too late to measure the optical peak.
 
Broadly speaking, there are two possibilities for the origin of the
optical emission which we observe when the GRB first comes into view
of {\em Swift} UVOT. The optical emission could be produced by the
shocks which are generated as the ejecta are slowed by the external
medium, and/or the optical emission could be related to prompt
gamma-ray emission, which in turn is thought to be produced
internally to the outflow. It is possible that a combination of 
both possibilities could contribute
to the observed emission and that the dominant contributor could be
different at the time of the settling exposure in different GRBs.

\begin{figure*}
\begin{center}
\includegraphics[width=170mm, angle=0]{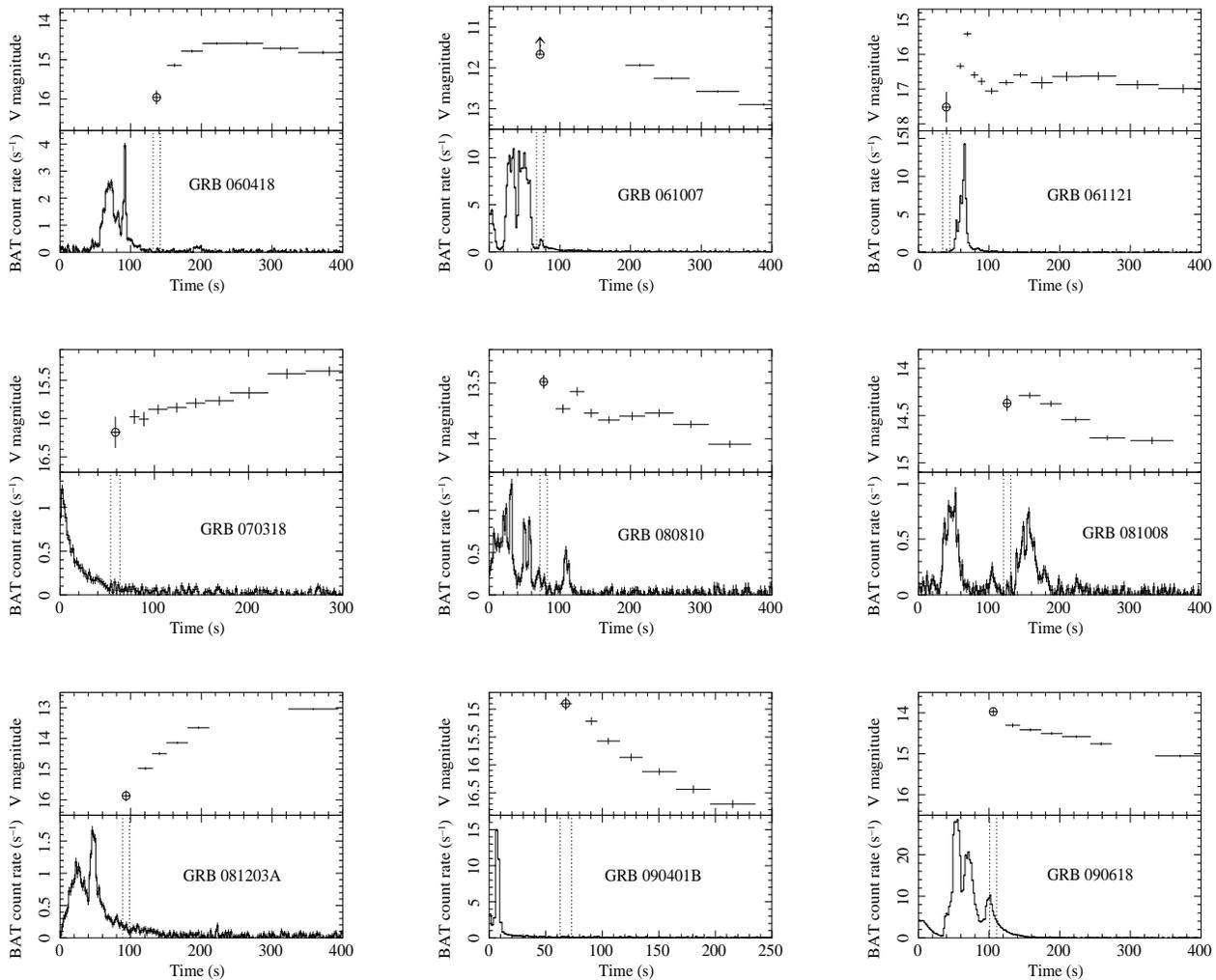}
\caption{Optical and gamma-ray lightcurves of the 9 GRBs for which T90 extends past the beginning of the UVOT settling exposure. For each GRB the top panel shows the optical lightcurve derived from UVOT and translated into the $V$ band as described in \citet{oates12}. The settling exposure is indicated with a circle. The bottom panel shows the full-band 15-350 keV lightcurve from the BAT. The dotted lines indicate the time range of the UVOT settling exposure.}
\label{fig:lightcurves}
\end{center}
\end{figure*}

We start by examining the possibility that the earliest optical
emission comes from the internal emission. Of our sample of 23 GRBs,
there are 14 GRBs for which the settling data were taken after the
conclusion of T90; for these GRBs we can assume that the earliest
optical emission is observed too late to be attributed to internal
emission. The remaining 9 GRBs, for which T90 encompasses the settling
exposure, are now examined in more detail. Fig. \ref{fig:lightcurves}
shows the UVOT and BAT lightcurves for these 9 GRBs. For GRBs 060418,
070318, 081203A, and 090401B, the settling exposure matches well the
extrapolation of data from later times, and appears to have little
correspondence to the prompt gamma-ray emission. In GRB\,090618 the
settling exposure shows enhanced optical emission compared to the
later time optical lightcure, and corresponds approximately to one of
the peaks in the prompt emission. Therefore the prompt emission could
plausibly contribute to the optical flux at the time of the settling
exposure for this GRB. In GRB\,061007 the settling exposure is
consistent with the extrapolation of the UVOT lightcure, 
but the UVOT
measurement is beyond the bright coincidence-loss limit, and hence
we have only a lower limit to the optical flux; 
the
settling exposure coincides with one of the peaks in the prompt
emission, so for GRB\,061007 we also consider it possible that the
prompt emission contributes to the optical flux measured in the
settling exposure.  In GRBs 061121, 080810 and 081008 the settling
exposures correspond to times of little or no gamma-ray emission, but
are followed by significant pulses of gamma rays and peaks in the
optical that suggest that the optical emission at the time of the
settling exposure and for some period subsequently might be related to
the prompt emission. Thus there are five GRBs, 061007, 061121, 080810, 081008
and 090618, for which the prompt emission might be contributing to the
optical flux at the time of the settling exposure, and/or at the time of
the peak optical flux.

We consider next the scenario in which the earliest optical emission
comes from the external shock, which is the standard paradigm for the
afterglow emission. 
Either the forward shock propagating into the external medium, or the reverse shock propagating back into the ejecta \citep{meszaros93} are viable mechanisms for producing the early optical emission.
Emission from an external shock at an observed
time $T_{\rm settle}$ after the beginning of the GRB implies that the
external shock is producing detectable optical emission at a distance
$R_{\rm settle}$ from the explosion site \citep{zhang04}:
\begin{equation}
R_{\rm settle} = 2 \Gamma^{2}\,c\,T_{\rm settle}\,(1+z)^{-1}
\end{equation} 
where $c$ is the speed of light, $z$ is the redshift of the GRB and
$\Gamma$ is the Lorentz factor of the relativistic outflow.  For a
characteristic $\Gamma=300$ and the observed range of 10~s~$<
T_{\rm settle}/(1+z) <$~70~s, $R_{\rm settle}$ ranges from $5.4\times 10^{16}$
and $3.8\times 10^{17}$~cm. For the majority of the GRBs in our sample
the optical lightcurve is already declining at the time of the
settling observation, so $R_{\rm settle}$ represents an upper limit to the
deceleration radius.

In Fig.~\ref{fig:tsettle_vs_t90} it is notable that four of the five GRBs
for which a robust optical peak is detected have $T_{\rm settle}<T_{90}$,
whereas the majority of the GRBs which are declining in the optical
from the first observation have $T_{\rm settle}>T_{90}$. To evaluate the
significance of the apparent trend for a larger proportion of GRBs
with optical peaks to have a $T_{\rm settle}<T_{90}$ than those without,
we apply Fisher's exact test,
 obtaining a $p$-value of 5.6 per cent for the null hypothesis that there 
is no trend. 
The trend is therefore of marginal (approximately
2$\sigma$) significance. However, it is interesting to note that two
of the GRBs in our sample have observations from ground based
observatories which caught the peak of the optical emission:
GRB~061007 with ROTSE \citep{rykoff09} and GRB~080319B with TORTORA
\citep{racusin08}. In both cases, the ground-based observations began
before the end of $T_{90}$; if these observations were substituted for
those of UVOT in Fig.~\ref{fig:tsettle_vs_t90} 
the Fisher's exact test would instead yield a $p$-value of just 1.8~per cent.

Naturally, it would be expected that the earlier the observations
begin, the greater the likelihood of observing the peak in the optical
emission, irrespective of $T_{90}$, so it is worth investigating
whether this is responsible for the tendency to observe the peak if
$T_{\rm settle}<T_{90}$. A Komogorov Smirnov (KS) test on the
distributions of $T_{\rm settle}/(1+z)$ for GRBs with and without optical
peaks results in
a $p$-value of 99.6 per cent for the null hypothesis that the two 
distributions are the same, hence the two distributions are indistinguishable.
 In contrast, a KS
test on the distributions of $T_{90}/(1+z)$ for the two groups of GRBs gives
a 
$p$-value of just 4.8 per cent.
This suggests that
within our sample, optical peaks are observed preferentially 
in GRBs with long $T_{90}$,
rather than in GRBs with early optical observations.
 
Theoretically, the timing of the peak in the optical lightcurve with
respect to $T_{90}$ depends on whether the afterglow is produced by
the collision of a thin or thick shell with the surrounding
interstellar medium. Specificially, the thick shell case corresponds
to $\Delta R > (E_{K}/nm_{p}\,c^{2})^{1/3}\Gamma_{0}^{-8/3}$ where
$\Delta R$ is the thickness of the shell, $E_{K}$ is the kinetic energy
of the shell, $\Gamma_{0}$ is the initial Lorentz factor of the shell,
$n$ is the density of the surrounding medium, $c$ is the speed of
light in vacuum and $m_{p}$ is the mass of a proton \citep{sari99}. It
is common to assume that the duration of the burst, $T_{90}$ is
approximately $\Delta R / c$, with the implication that for thick shells
the peak time of the optical emission should be comparable to $T_{90}$
\citep{sari99,kobayashi00}.
A peak time for the optical emission which is significantly delayed
with respect to the end of $T_{90}$ is therefore usually taken to
imply the thin shell case \citep{molinari07, rykoff09}. The numerical
modelling of \citet{kobayashizhang07} would suggest that
$T_{\rm peak}/T_{90}<2$ may be a realistic expectation for a thick
shell. Thick shells should also exhibit relatively shallow afterglow
rise profiles, with $\alpha< 1$ where $L\propto t^{\alpha}$
\citep{kobayashi00,kobayashizhang07}.  For the afterglows with well
defined peaks, $\alpha$ can be estimated from the timing and magnitude
of the peak relative to the settling exposure:
\begin{equation}
\alpha=\frac{M_{\rm settle}-M_{\rm peak}}{2.5(\log_{10}T_{\rm peak}-\log_{10}T_{\rm settle})}
\label{eq:rise}
\end{equation}
For the five gamma-ray bursts in our sample for which a secure peak
has been observed in the UVOT lightcurve, the values of
$T_{\rm peak}/T_{90}$ and $\alpha$ for the rise phase are given in
Table~\ref{tab:peaks}. None of the GRBs satisfy both
$T_{\rm peak}/T_{90}<2$ and $\alpha<1$, implying that all
of them have thin shell afterglows. Consequently, if there is a
connection between $T_{90}$ and the observability of the optical peak,
it is not because the outflows are thick shells.

In the thin shell regime, the peak of the optical emission is expected
to occur when the shell reaches the deceleration radius, at which the
shell has swept up a mass of material from the surrounding
interstellar medium equal to $\sim 1/\Gamma$ times the initial rest mass of
the shell \citep{zhang04}. In this case, the timing of the afterglow 
peak relative to
the start of the GRB should be independent of $T_{90}$, and therefore a
connection between $T_{90}$ and the observability of the optical peak
emission is not expected in the thin shell case.

An alternative explanation for the trend for GRBs with large $T_{90}$
to present well-defined peaks in their optical emission would be for
the prompt emission to be contributing to the optical peak. As we have
already discussed, the prompt emission is unlikely to contribute to
the observed optical emission except, perhaps, in a small minority of the GRBs
in our sample. Of those with well-defined optical peaks, only
GRB\,061121 \citep[see also ][]{page07} is a good candidate for a prompt 
contribution to the optical peak. 
Nonetheless, the trend relating $T_{90}$ to detection of
optical peaks is based on such a small number of sources, that it
could be explained as a combination of prompt optical emission in
GRB\,061121 together with a statistical fluctuation in these
properties amongst the remaining GRBs.

\begin{table}
\caption{Parameters for GRBs with well-defined peaks in their optical lightcurves. $T_{\rm peak}$ is the mid-time of the brightest optical measurement, and $\alpha$ is the power-law index of the optical rise, as defined by Equation \ref{eq:rise}.
}
\label{tab:peaks}
\begin{tabular}{lccc}
GRB         &$T_{\rm peak}$&$T_{\rm peak}/T_{90}$&$\alpha$\\
            &(s)      &                &        \\
\hline
&&&\\
GRB\,060418 &264.5    &  1.84          &$ 1.92\pm 0.24$\\    
GRB\,060607 &207.3    &  3.42          &$ 2.83\pm 0.35$\\    
GRB\,061121 &69.5     &  0.85          &$ 3.44\pm 0.71$\\    
GRB\,070318 &340.8    &  3.15          &$ 0.44\pm 0.11$\\    
GRB\,081203A&358.4    &  1.62          &$ 1.94\pm 0.12$\\    
&&&\\
\hline
\end{tabular}
\end{table}

\section{Conclusions}
\label{sec:conclusions}

We have examined the optical emission from a sample of
optically-bright (peak $V<16$ mag) GRB afterglows as they first came
into view with {\em Swift} UVOT. 
In one case (GRB\,061007) the first second of the settling exposure was excluded because the spacecraft attitude reconstruction was not good enough to prevent trailing of stars at the beginning of the exposure. In another case (GRB\,091020) the first second of the settling exposure was excluded because we found evidence that the UVOT photocathode voltage is still ramping up during that first second. 
The photometric quality of the UVOT
settling exposures was verified using photometry of bright stars in
the field of view. 
These stars were found to be of very similar brightness 
in the settling exposures
to subsequent UVOT images (differing on average by $0.021\pm 0.009$ mag). 
The settling exposures are
therefore considered to be good enough to derive photometry for the
GRB afterglows.

Of the sample of 23 GRBs, all are detected in the UVOT settling
exposures, and hence in every case the optical emission had already
begun by the time of the settling exposure. In 9 of the GRBs, the
settling exposure took place within $T_{90}$ of the prompt gamma-ray
emission. Five GRBs have well defined optical peaks, with measured
rises of $>0.5$~mag in their optical lightcurves following the
settling exposure. A trend is found, with marginal statistical
signficance (2$\sigma$), for these GRBs with well-defined optical
peaks to have large values of $T_{90}$, and to be observed before the
conclusion of $T_{90}$. Such a trend would be expected from
thick-shell afterglows, but the timing of their optical peaks and the
temporal indices of their optical rises rule out thick-shell
behaviour. Instead, a contribution from the prompt emission to the
optical peak in one or more GRBs could account for the trend.

\section*{Acknowledgments}
\label{sec:acknowledgments}

This work has been supported by the UK Space Agency under grant
ST/P002323/1 and the UK Science and Technology Facilities Council
under grant ST/N00811/1. This work has made use of the UK {\em Swift}
Science Data Centre, hosted at the University of Leicester, UK. SRO gratefully acknowledges the support of a Leverhulme Trust Early Career Fellowship.

\bibliographystyle{mn2e}

\appendix

\vspace{3mm}
\noindent
{\bf APPENDIX A: PHOTOMETRIC PERFORMANCE OF THE UVOT DURING THE SETTLING EXPOSURES}
\vspace{1mm}

Settling exposures are taken at the final stage of the {\em Swift}
spacecraft slew, when the target has entered the field of view of
UVOT, but the spacecraft is still moving.
Bright stars observed simultaneously with the GRB afterglows were used
to verify the photometric validity of the UVOT photometry used in this
study. Source measurements were made using {\sc uvotsource} with a
five arcsec radius aperture and background measurements were made
using the same larger apertures that were employed in the afterglow
photometry (Section \ref{sec:observations}). Two measurements were made
for each star. The first is from the image formed from the settling
data that was used for afterglow photometry. The second measurement
was made using the sum of the other $V$-band images obtained during
the same observation sequence of the GRB afterglow, excluding any
images exhibiting attitude problems. Photometry obtained from the
second measurement is invariably of higher precision than that
obtained from the settling data because the exposure times are much
longer, and as the photocathode voltage is stable after the settling
exposure the photometry is reliable. Table A1 gives the photometry for 
the stars in each GRB field of view.

\begin{figure*}
\begin{center}
  \includegraphics[width=150mm, angle=0]{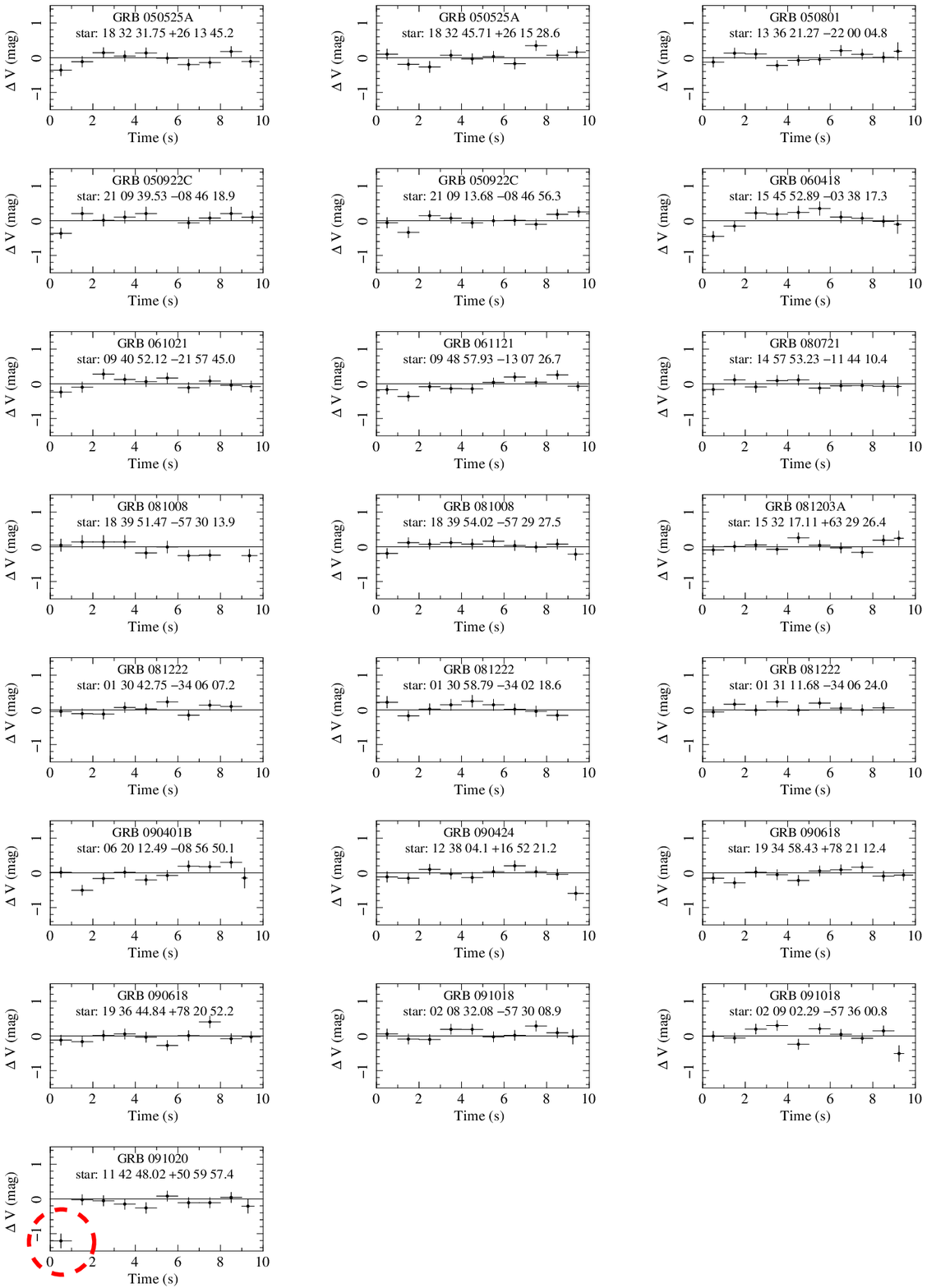}
\end{center}

\noindent {\bf Figure A1} Lightcurves of comparison stars brighter
than $V=13$~mag during the settling exposures.  $\Delta V$ is computed
as $V_{\rm settled} - V$ where $V_{\rm settled}$ is the $V$ magnitude
in the subsequent settled exposures, so that negative values
correspond to the reference star appearing fainter in the settling
exposure. Note the large, negative $\Delta V$ during the first time
bin for the GRB\,091020 comparison star~11~42~48.02~+50~59~57.4
(indicated by the dashed circle in the bottom left panel), suggesting
that the voltage ramp-up of the photocathode took place during the
first second of the settling exposure in this case. On two occasions,
once each for the two brightest comparison stars (GRB\,050922C,
star~21~09~39.53~$-$08~46 18.9 and GRB\,081008,
star~18~39~51.47~--57~30~13.9) positive fluctuations push the count
rate above the bright calibration limit for coincidence loss
\citep{poole08, page13}. These two datapoints have been omitted from
the relevant lightcurves.
\end{figure*}

\begin{table*}
{\bf Table A1.} Photometry of bright stars within the UVOT fields of view 
for the GRB afterglows used in the study. $\Delta V$ is computed as $V_{\rm settled}-V_{\rm settling}$ where $V_{\rm settling}$ is the $V$ magnitude in the settling exposure and $V_{\rm settled}$ is the $V$ magnitude in the subsequent settled exposures, so that negative values correspond to the reference star appearing fainter in the settling image.
\begin{tabular}{cccccc}
&&&&&\\
GRB&\multicolumn{2}{c}{Star}     &$V$ mag in & $V$ mag in  & $\Delta V$\\
   &             RA&dec          &settling   & subsequent  & (mag)     \\
   &      (J2000)  &(J2000)      & data      &  images     &           \\   
\hline
GRB\,050525A&18 32 31.75&$+$26 13 45.2&$12.95\pm0.05$&$12.89\pm0.02$&$-0.06\pm0.06$\\
$"$    &18 32 45.71&$+$26 15 28.6&$12.90\pm0.05$&$12.91\pm0.02$&$+0.01\pm0.06$\\
&&&&&\\
GRB\,050801 &13 36 21.27&$-$22 00 04.8&$12.26\pm0.05$&$12.27\pm0.02$&$+0.01\pm0.06$\\
$"$    &13 36 23.93&$-$21 59 04.2&$14.81\pm0.10$&$14.82\pm0.02$&$+0.01\pm0.10$\\
$"$    &13 36 40.88&$-$21 55 58.7&$14.99\pm0.11$&$14.85\pm0.02$&$-0.14\pm0.11$\\
&&&&&\\
GRB\,050922C&21 09 39.53&$-$08 46 18.9&$11.71\pm0.06$&$11.77\pm0.02$&$+0.06\pm0.06$\\
$"$    &21 09 13.68&$-$08 46 56.3&$12.74\pm0.05$&$12.75\pm0.02$&$+0.01\pm0.05$\\
$"$    &21 09 36.04&$-$08 42 47.8&$14.04\pm0.07$&$14.00\pm0.02$&$-0.04\pm0.07$\\
&&&&&\\
GRB\,060418 &15 45 52.89&$-$03 38 17.3&$11.85\pm0.06$&$11.85\pm0.02$&$+0.00\pm0.06$\\
$"$    &15 45 56.35&$-$03 36 26.3&$14.23\pm0.08$&$14.17\pm0.03$&$-0.05\pm0.08$\\
$"$    &15 45 48.90&$-$03 40 17.8&$14.19\pm0.08$&$14.20\pm0.03$&$+0.01\pm0.08$\\
&&&&&\\
GRB\,060607 &21 58 45.54&$-$22 30 47.9&$13.78\pm0.07$&$13.83\pm0.02$&$+0.04\pm0.07$\\
$"$    &21 58 53.31&$-$22 26 31.4&$14.14\pm0.08$&$14.24\pm0.02$&$+0.11\pm0.08$\\
&&&&&\\
GRB\,060908 &02 07 10.34&$+$00 23 16.6&$13.26\pm0.06$&$13.15\pm0.02$&$-0.11\pm0.06$\\
$"$    &02 07 20.98&$+$00 18 44.8&$14.44\pm0.09$&$14.36\pm0.02$&$-0.09\pm0.09$\\
$"$    &02 07 23.32&$+$00 20 39.1&$15.82\pm0.16$&$15.80\pm0.03$&$-0.02\pm0.16$\\
&&&&&\\
GRB\,061007 &03 05 01.35&$-$50 28 19.6&$12.60\pm0.06$&$12.63\pm0.02$&$+0.03\pm0.06$\\
$"$    &03 05 40.60&$-$50 31 35.0&$13.94\pm0.08$&$13.82\pm0.02$&$-0.12\pm0.08$\\
&&&&&\\
GRB\,061021 &09 40 52.12&$-$21 57 45.0&$12.68\pm0.05$&$12.69\pm0.02$&$+0.01\pm0.06$\\
$"$    &09 40 48.14&$-$21 55 42.2&$13.51\pm0.06$&$13.52\pm0.02$&$+0.01\pm0.07$\\
$"$    &09 40 45.48&$-$22 00 13.9&$14.35\pm0.08$&$14.14\pm0.02$&$-0.21\pm0.08$\\
&&&&&\\
GRB\,061121 &09 48 57.93&$-$13 07 26.7&$12.41\pm0.05$&$12.35\pm0.02$&$-0.06\pm0.05$\\
$"$    &09 49 03.24&$-$13 09 28.4&$14.42\pm0.08$&$14.40\pm0.02$&$-0.02\pm0.08$\\
$"$    &09 49 04.68&$-$13 10 31.0&$14.63\pm0.09$&$14.61\pm0.02$&$-0.02\pm0.09$\\
&&&&&\\
GRB\,070318 &03 13 55.40&$-$42 53 14.1&$15.63\pm0.15$&$15.53\pm0.02$&$-0.10\pm0.15$\\
$"$    &03 14 10.25&$-$42 55 32.5&$15.53\pm0.14$&$15.56\pm0.02$&$+0.03\pm0.14$\\
$"$    &03 13 59.07&$-$42 57 08.4&$15.88\pm0.17$&$15.79\pm0.02$&$-0.09\pm0.17$\\
&&&&&\\
GRB\,080721 &14 57 53.23&$-$11 44 10.4&$13.00\pm0.06$&$12.98\pm0.02$&$-0.03\pm0.06$\\
$"$    &14 58 04.47&$-$11 41 22.6&$15.04\pm0.11$&$14.88\pm0.03$&$-0.16\pm0.11$\\
&&&&&\\
GRB\,080810 &23 47 05.23&$+$00 15 45.2&$14.02\pm0.07$&$14.03\pm0.03$&$+0.00\pm0.08$\\
$"$    &23 46 58.49&$+$00 18 30.7&$14.63\pm0.09$&$14.55\pm0.03$&$-0.08\pm0.09$\\
$"$    &23 47 02.82&$+$00 22 06.9&$14.64\pm0.09$&$14.71\pm0.03$&$+0.07\pm0.09$\\
&&&&&\\
GRB\,081008 &18 39 51.47&$-$57 30 13.9&$11.70\pm0.06$&$11.66\pm0.03$&$-0.05\pm0.06$\\
$"$    &18 39 54.02&$-$57 29 27.5&$12.30\pm0.05$&$12.33\pm0.02$&$+0.02\pm0.06$\\
$"$    &18 40 00.95&$-$57 27 17.5&$14.22\pm0.08$&$14.24\pm0.03$&$+0.03\pm0.08$\\
&&&&&\\
GRB\,081203A&15 32 17.11&$+$63 29 26.4&$12.60\pm0.05$&$12.63\pm0.02$&$+0.03\pm0.06$\\
$"$    &15 31 51.63&$+$63 29 47.6&$13.70\pm0.07$&$13.61\pm0.03$&$-0.10\pm0.07$\\
$"$    &15 32 27.45&$+$63 32 34.2&$15.26\pm0.12$&$14.88\pm0.03$&$-0.38\pm0.13$\\
&&&&&\\
GRB\,081222 &01 30 58.79&$-$34 02 18.6&$11.89\pm0.06$&$11.92\pm0.02$&$+0.03\pm0.06$\\
$"$    &01 30 42.75&$-$34 06 07.2&$12.47\pm0.05$&$12.47\pm0.02$&$+0.00\pm0.06$\\
$"$    &01 31 11.68&$-$34 06 24.0&$12.70\pm0.05$&$12.78\pm0.02$&$+0.08\pm0.06$\\
&&&&&\\
GRB\,090401B&06 20 12.49&$-$08 56 50.1&$12.13\pm0.05$&$12.08\pm0.02$&$-0.05\pm0.06$\\
$"$    &06 20 21.95&$-$08 59 30.1&$13.66\pm0.07$&$13.58\pm0.02$&$-0.07\pm0.07$\\
$"$    &06 20 16.32&$-$08 59 48.6&$13.62\pm0.07$&$13.61\pm0.02$&$-0.01\pm0.07$\\
\hline
\end{tabular}
\end{table*}

\begin{table*}
{\bf Table A1.} continued\ \ \ \ \ \ \ \ \ \ \ \ \ \ \ \ \ \ \ \ \ \ \ \ \ \ \ \ \ \ \ \ \ \ \ \ \ \ \ \ \ \ \ \ \ \ \ \ \ \ \ \ \ \ \ \ \ \ \ \ \ \ 
\begin{tabular}{cccccc}
&&&&&\\
GRB&\multicolumn{2}{c}{Star}&$V$ mag in & $V$ mag in  & $\Delta V$\\
   &          RA&dec        &settling   & subsequent  & (mag)     \\
   &     (J2000)&(J2000)    & data      &  images     &           \\   
\hline
&&&&&\\
GRB\,090424 &12 38 04.01        &+16 52 21.2           &$12.84\pm0.05$&$12.78\pm0.03$&$-0.06\pm0.06$\\
$"$    &12 38 08.61        &+16 50 23.9           &$14.15\pm0.08$&$14.10\pm0.03$&$-0.05\pm0.07$\\
$"$    &12 37 50.82        &+16 50 11.7           &$14.10\pm0.08$&$14.18\pm0.03$&$+0.08\pm0.08$\\
&&&&&\\
GRB\,090618 &19 36 44.84        &+78 20 52.2           &$12.17\pm0.05$&$12.11\pm0.02$&$-0.05\pm0.06$\\
$"$    &19 34 58.43        &+78 21 12.4           &$12.62\pm0.05$&$12.56\pm0.02$&$-0.06\pm0.06$\\
$"$    &19 36 34.28        &+78 19 20.7           &$14.27\pm0.08$&$14.16\pm0.02$&$-0.11\pm0.08$\\
&&&&&\\
GRB\,090812 &23 32 36.14        &-10 39 02.2           &$14.39\pm0.08$&$14.28\pm0.04$&$-0.11\pm0.09$\\
$"$    &23 33 01.04        &-10 37 06.4           &$15.35\pm0.13$&$15.14\pm0.05$&$-0.21\pm0.14$\\
$"$    &23 32 40.95        &-10 36 11.2           &$16.00\pm0.18$&$15.98\pm0.07$&$-0.01\pm0.19$\\
&&&&&\\
GRB\,091018 &02 08 32.08        &-57 30 08.9           &$12.22\pm0.05$&$12.29\pm0.02$&$+0.06\pm0.06$\\
$"$    &02 09 02.29        &-57 36 00.8           &$12.51\pm0.05$&$12.53\pm0.02$&$+0.02\pm0.06$\\
$"$    &02 08 50.91        &-57 33 58.0           &$14.43\pm0.09$&$14.46\pm0.02$&$+0.03\pm0.09$\\
&&&&&\\
GRB\,091020 &11 42 48.02        &+50 59 57.4           &$12.88\pm0.05$&$12.68\pm0.02$&$-0.20\pm0.06$\\
$"$    &11 43 07.16        &+50 57 28.8           &$15.21\pm0.12$&$15.17\pm0.02$&$-0.04\pm0.12$\\
$"$    &11 42 47.24        &+50 57 24.2           &$15.49\pm0.14$&$15.59\pm0.02$&$+0.10\pm0.14$\\
&&&&&\\
GRB\,100906A&01 54 31.65        &+55 40 14.7           &$13.29\pm0.06$&$13.27\pm0.02$&$-0.02\pm0.06$\\
$"$    &01 54 38.13        &+55 37 14.5           &$13.31\pm0.06$&$13.31\pm0.02$&$+0.00\pm0.06$\\
$"$    &01 55 00.63        &+55 38 55.3           &$13.66\pm0.07$&$13.62\pm0.02$&$-0.05\pm0.07$\\
\hline
\end{tabular}
\end{table*}

We can further examine the photometric performance of the UVOT during
the settling exposures by dividing the settling exposure into smaller
time bins. In this way, we can investigate whether the photocathode
voltage ramp-up is complete, and the detector is stable when the
settling exposure begins, or whether the detector performance is
changing during the settling exposure. Due to the small integration
times involved, this is only practical for the brighter comparison
stars. Therefore, for each of the comparison stars in Table A1
brighter than $V=13$ mag (except for those in the field of
GRB\,061007)\footnote{As noted in Section \ref{sec:observations},
  there is a problem with the spacecraft attitude reconstruction in
  the first second of the GRB\,061007 settling exposure, and so this
  is not an appropriate dataset for investigation of the photometric
  performance at the beginning of the settling exposure.}, we have
generated lightcurves with a cadence of 1~s. The lightcurves are shown
in Fig.~A1.

\label{lastpage}

\end{document}